# From Teaching to Coaching: A Case Study of a Technical Communication Course

**Rayed Alghamdi, Seyed M. Buhari & Madini O. Alassafi**

*Department of Information Technology*
*Faculty of Computing and Information Technology*
*King Abdulaziz University, Jeddah, Saudi Arabia*
*{raalghamdi8; mesbukary; malasafi} @kau.edu.sa*

**Abstract**

One of the leading university goals is to provide the students with the necessary skills for better functioning in their future studies. Gaining and developing skills, both technical and soft skills, are the critical building blocks for a successful career. The traditional teaching process, which includes delivering lectures and conducting exams, emphasizes the upliftment of technical knowledge rather than building self-esteem and enhancing skills development among students. Development of non-technical skills like self-esteem, life-long learning among students is vital for a successful career. This paper targets to achieve the objective of identifying ways to empower students with non-technical skills along with technical skills. The approach adopted in this research work is to transform the delivery of a faculty-wide course from teaching to coaching. One salient difference between teaching and coaching is to move students motivation away from grades towards life-long learning. Thus, university students maximize skills through coaching like course structure, course delivery, course assessment, and student involvement. As a case study, the proposed approach was successfully tested and validated on a course taught in the Faculty of Computing and Information Technology (FCIT) at King Abdulaziz University (KAU). The outcome of this case study reveals that there is substantial potential towards enhancement of self-responsibility. In the future, this approach can be extended for other courses that aim to develop skills such as English language courses, computer, and communication skills courses in the preparatory/foundation year.

## 1 INTRODUCTION

In the FCIT curriculum, students must complete 200 hours of real workplace training. This training should be conducted as a college requirement for almost one year before graduation [2]. Since 2017, we have noticed a shortcoming in some necessary soft skills for the summer training students in the information technology (IT) department of FCIT. This might be due to the students' adaptability towards the workplace environment, which might be different from academia. However, the IT department is entitled to provide the necessary soft skills to the students. A study was conducted to foster IT students internship program. In that study, the skills gap between IT internship students and leaders in the Saudi Arabian IT industry, both public and private, were explored. Forty-seven skills were evaluated from two perspectives: students and the industry. The study highlighted the importance of soft skills. The top five ranked skills were communication skills (oral and written), ability to work in teams, willingness to learn new skills, and flexibility/adaptability [3]. The outcome of this study was compared with our current IT curriculum [4]. This urged us to conduct a substantial review of how efficiently our current curriculum, which mainly targets enhancing technical skills, can be adapted to enhance soft skills. The review was conducted under the umbrella of the latest report of information technology curricula of the Association for Computing Machinery (ACM) and the IEEE Computer Society [5]. One of the major changes that aimed to target our students' soft skills was investigated in the Technical Communication course. This course is taught for first-year students at FCIT, irrespective of their major, as it is the basis of delivering the message for everyone. Over the past years, this course has been taught traditionally by delivering lectures, submitting assignments, and conducting two or three exams. Since the start of the current year (2019), a decision was taken with the full support of the IT department to transform from a traditional teaching process to a coaching approach, supported using blackboard. The course content on the blackboard, see Figure 1, was designed based on the FORCE strategy guidelines [6].

In the following sections, we start by providing pedagogical background on the learning process, followed by explaining our developed approach in coaching the Technical Communication course, and conclude with giving hints on how this approach can be extended for other courses in order to maximize the skills development for the preparatory-year students.

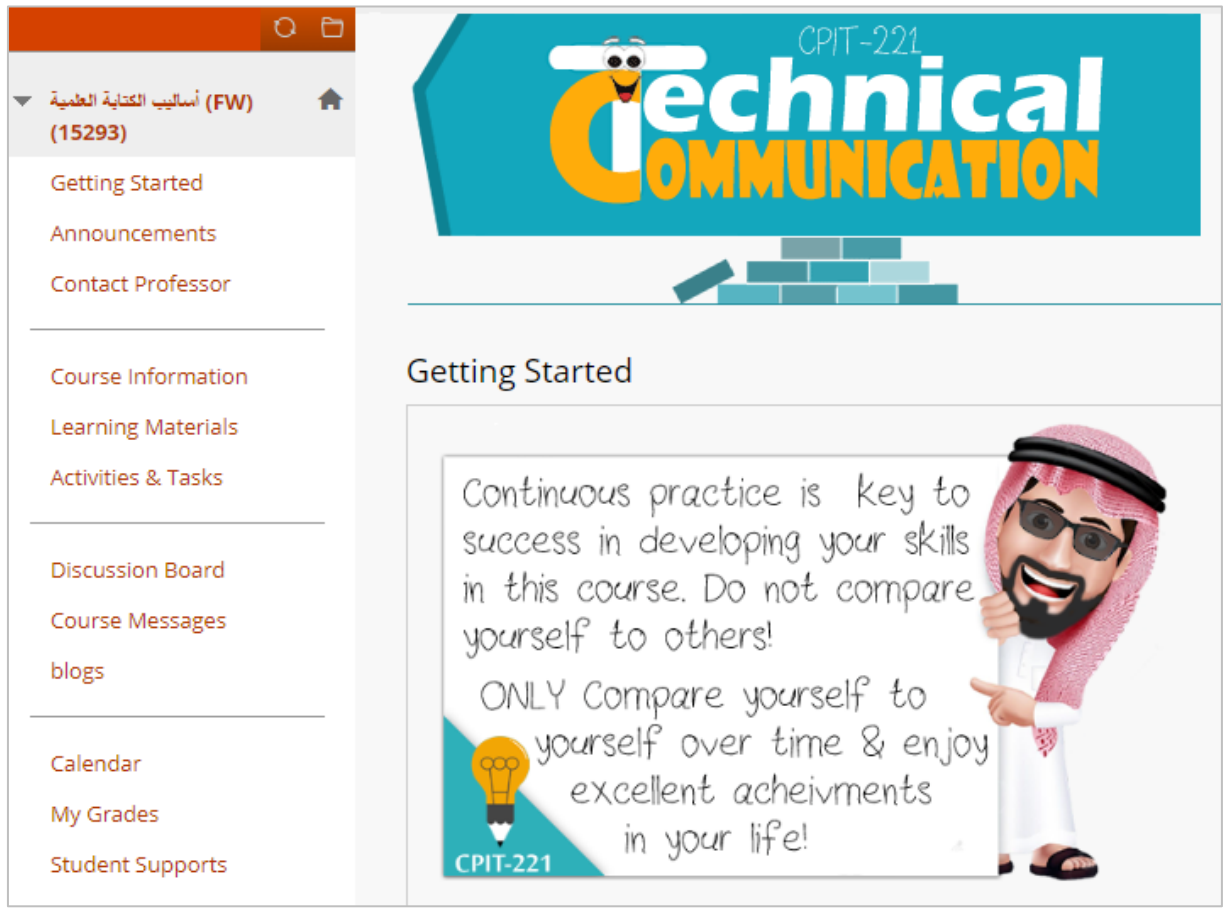

*Fig. 1: Main page of the Technical Communication course on the blackboard.*

## 2   PEDAGOGICAL BACKGROUND

Enhancing the skills of the students in a curriculum could be addressed through (1) approach of course content delivery, (2) course assessment strategies, (3) ordering of course content, or framing a syllabus.

Courses are framed mainly with technical concerns in the mindset of the instructor. This is the typical nature of academicians who are in the field of computing or information technology. While framing and delivering the courses, the instructor's concern is about covering the material and evaluating the students from their perspective [7]. This motivates the instructors to add as many varied assessments as possible. In a course, along with exams, there are many assignments and projects assigned to students. Such an approach could hinder the progress of the student when the course is skills-based [8]. This is because students, as like the instructor, will be more concerned about the assessments from the perspective of grades. Thus, they will always tend to think of what is covered in the assessments and how to attain success in them instead of building the necessary skills for the workplace [9]. It is also visible from the end of semester improvement suggestions that instructors tend to add more assessments whenever the students' performance is poor.

For example, in a course like Technical Communication, students must build technical communication skills, which include writing, reading, speaking, and presenting. The course is also targeted at developing skills like the ability to work in teams, learn new skills, and be flexible/adaptable. These could be enhanced only with sufficient effort and time [10]. Enhancement of one's skills will depend on the amount of time allocated for the skills development, with the objective to enhance the respective skill and not be distracted by the grades or assessments behind them [9].

The challenge here is how to motivate the students who look towards assessments to give due time for a task, towards their skills development [9]. This is where coaching comes ahead of teaching. Tasks assigned to the students should consider various factors like tasks from other courses, state-of-mind of the students, etc. For example, during the exam periods of the semester, students tend to be overwhelmed with various exams and tend to give lighter importance for other tasks [11]. As could always be seen, students who are active in sports tend to reduce them during examination periods.

While teaching a course, instructors are allocated respective sections and teach the course independently. This provides students with learning a course from the perspective of an instructor. In some countries, university students are asked to take exams prepared by an instructor, unknown to them [12]. Thus, they are preparing themselves for any question that could arise from a given syllabus. Generally, the outcome of such an exercise is that students score higher in the internal assessments than the external assessments [4]. The average of internal and external assessment is considered as the total score of the student. The ratio of the internal and external assessments towards the total score might be one of the deciding factors towards measuring the students' capabilities. For example, in educational systems, where the evaluation was changed from 100% external assessment to a mixture of internal and external assessments, a substantial increase in the students' marks is visible [12]. Furthermore, Saudi Arabia is also thinking on the Exit Exam lines for the graduating students [13]. Thus, a student is evaluated on the subject matter from different perspectives. Considering this aspect of teaching and learning, instructors of the Technical Communication course have opted to co-teach the sections. The course's certain skills were taught by a specific instructor based on his/her interest and strength.

The order of delivery of teaching materials in the Technical Communication course might affect the way the teaching or coaching is made to the students. From articles to sentence level, moving from top to bottom might be appropriate for teaching this course. But, to bring students from the base towards a better level, coaching must start from the primitive, the sentences or words level. Having learned some pre-year English and Communication courses, students are at the level of the words already and thus they should be coached from sentence level to article or report level. Our focus in this course is on weekly achievements to monitor the students' progress and guide them to develop their skills better. This is how a typical skills-based course should be designed as recommended by Kennedy [15]. It is by focusing on practice and achievement rather than on what has been taught or delivered. The following section details our experience on how coaching of the Technical Communication course was conducted.

## 3   METHODOLOGY

Based on the literature review, we planned to accomplish skills-development in our course through four approaches: (1) Course Structure, (2) Course Delivery, (3) Course Assessments, and (4) Student involvement.

*Course Structure* - Being a skills-based course, the course structure was aligned like that of training programs. The complete program or the course was structured on a weekly basis, instead of providing the students with chapters-based course delivery. Skills learned in each week acts as the foundation for the next week's materials. Each week is organized according to three main aspects: objectives, lecture notes, and activities. Each week focused on only two objectives: one targeting the knowledge and the other targeting the practice/activity levels. In contrast to reading and memorizing lecture materials, the emphasis is given to practicing various activities during the lecture period, under the instructors' guidance. Lecture notes are considered a reference for the materials. They motivate students to perform the assigned activities. Each week typically ends with activities and tasks; in-class and take-home. Figure 2 below presents a typical structure of a week's content on the blackboard.

*Course Delivery* - The course content was organized on the blackboard. Blackboard is a learning management system that has been adopted for the teaching and learning processes at KAU since 2014 [15]. It is rich with tools that allow instructors to manage delivering the material and monitor students/trainees' progress. In addition, blackboard provides an attractive content platform, with rich tools that enable teachers/coaches to track the students' progress. Thus, the blackboard is one of the key elements in delivering this course effectively.

One key feature available on the blackboard is Discussion Board. The discussion board provides an interactive area where students can only present their work but also they can present themselves to others. Since our course targets communication skills, using a discussion board to submit the students video recordings is one step forward to break their fear/reluctance talking to a camera or a group of people. Due to a lack of self-esteem, students find speaking in public as a hurdle. Such a hurdle could be crossed through social media and interaction within the classroom through the discussion board. Students are given the assurance that mistakes are most welcome and will never be criticized. In this way, students are motivated to share their work and benefit from each other's work and feedback.

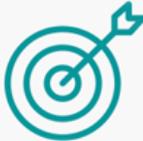

*Fig.2: The typical week's content structure of the Technical Communication course on the blackboard*

***Course Assessment*** - Peer assessment tool, available on the blackboard, was employed in this course effectively. This tool allows students to submit their work as well as peer review other students' works. This tool contributes to adding more people to the coaching process. It is not only a single-main person, i.e., teacher, who does everything for students. Students participate in the evaluation process, and they provide valuable feedback to each other. In addition, and most importantly, in these types of tasks, students learn not by submitting their works but also by evaluating and correcting others' mistakes; it looks like the common saying "catch two birds with one stone"!

***Student Involvement*** - Blackboard has tools that enable students to collaborate. Due to the way of students learning until the university level, they tend to be sole learners. This will hinder their progress once they enter the workplace. Thus, preparing the students for such an environment is necessary. Grouping of students for tasks is done in two ways: (1) Unfamiliar: randomized groups created between students from various sections of the course, (2) Familiar: students use the grouping tools on the blackboard to form their teams. These train the students to be adaptive and flexible when situations change in a direction that they may/may not like. Workplace expects you to work with whom you are either familiar or unfamiliar. In these tasks, students practice leadership and team functioning skills. One task of this course involves finding ways of doing business online and function the way that works to earn a minimum of 100 Saudi Riyals within one week, from the date of team formation. On the blackboard grouping pages, they have different tools that enable the coach/teacher to track each group's progress, such as blogs, journals, discussion boards, file exchanges, etc.

Finally, the adaptive release feature on the blackboard is handy in a skills-based course. As discussed earlier, the objective of this study is to move from teaching to coaching. In coaching, grades should not be the focus of both coaches and trainees. Grades will be gained based on the skills that were achieved. It is not about a one-time submission like an exam or an assignment where the gained grade is final. The trainees are given several attempts to complete a specific task, and they will not be allowed to move forward to further tasks unless the prior ones are satisfactory. This significantly assists the coach in adapting towards variances in knowledge and experiences among learners. Using one unified teaching/coaching method might be boring for some and difficult for others [16]. By using the adaptive method, learners can be fast/slow to complete the assigned tasks based on their knowledge and experiences. Every individual is given time and attempts that suit his/her needs.

This methodology is still evolving through our current pilot study. Preliminary results are discussed in the following section.

## 4  IMPLICATIONS AND APPLICATIONS

Improvements or changes are expected to be continuous and happen over a longer period. Even by the first quarter of the current semester, initial changes have already started to appear. Students' who were initially reluctant to communicate and find all sorts of excuses, especially in recording and sharing videos, overcame their initial hurdles and started to enhance their self-esteem. This change was made possible due to:
1. regular motivation of the students from the main course page on blackboard itself,
2. students are always asked to look for their own mistakes, as if they are looking at a mirror, instead of looking at others, and
3. instructors themselves initiating such a communication activity.

Self-responsibility is another important factor for a successful career. Students in the Technical Communication course are given the freedom to attend or not attend a lecture. Students have started to understand the importance of classroom-based interactive skills development. As the course is not delivered as a lecture, students are self-motivated and attend the lectures even without any attendance taken during the lecture period. It is not only the attendance that has been noted but also the interactions in the class have considerably improved. Thus, enhanced responsibility, which is one of the core aims of this course, is remarkably achieved.

Self-responsibility enhances further through self-motivation. Students are motivated to enhance their writing skills, by learning from their own mistakes. Students do their weekly writing tasks through this self-motivation, even after knowing that this task is not graded. While gaining experience through their own mistakes, students were given previously completed student reports to identify as well as correct potential errors in styles of technical writing. Students themselves were concerned about the quality of the reports they reviewed. This was used as a foundation to indicate that these poorly written reports resulted from students who kept thinking of grades instead of developing their skills. Since then, we keep noticing an excellent improvement in the students' attitude and self-motivation.

Workplace environment expects professionals to work on projects, as groups and not as an individual. A survey conducted among Technical Communication course students during the beginning of the semester reveals that almost 80% of the students prefer to work on course assignments or tasks as an individual. This kind of student mindset needs transformation towards a professional mindset. Materials related to teamwork and collaboration were covered in the Technical Communication course. During these lectures, students were positively encouraged towards group work while allowing them to discuss the cause for disinterest in group work. Students of this modern era tend to prefer online games than physical games. Taking an example from online gaming, it was indicated that developing a game requires diversified skills that are appropriately contributed towards the application development by different development team members. Thus, it is clearly stated that as future software professionals, they have to prepare themselves towards the workplace environment, which strongly relies on teamwork.

All the indicators mentioned above of our strategy's success are results achieved through the support of co-teaching by instructors, salient features of blackboard, and the support and cooperation from students. The same approach can be applicable and extendable to other courses that focus on developing skills.

## 5  RECOMMENDATIONS

We are highly motivated to extend and share the benefit of our experience. One of the most important times for university students is the preparatory year as it is the bridge between pre-university and university life. The preparatory year aims to empower students with the necessary skills to perform better in their next four studying years. It is about skills, so we believe that adopting our coaching strategy could further enhance and maximize the students' skills. There are important skills-based courses like English and Arabic languages, communication, computer, and mathematical skills. By looking at the dedicated teaching hours for every individual course, two or three hours a week will not help the students at skills/action/practice level. For this reason, we suggest adopting our strategy that is empowered by the effective usage of the blackboard. So, instructors and students spend more time to develop their skills by doing and practicing.

# 6 CONCLUSION

Understanding the nature of university students and their state-of-mind is a challenge by itself. Keeping this challenge in mind, instructors need to broaden their views of delivering skills-based courses. Expecting that marks gained will reflect the actual skills development by the students might be fruitful with self-motivated students. For the rest, an extra mile effort is needed to achieve self-esteem and success among students. This paper took such a task in the Technical Communication course, which is a first-year course in FCIT faculty. Adhering to the mode of coaching instead of teaching with an emphasis on self-esteem and moving the students from grades-specific thinking has helped everyone involved. Preliminary results reveal that we are currently able to see the outcome of the efforts or changes made. Further study will be made at the end of the semester and will be reported in due time.

## ACKNOWLEDGMENT

The authors express their great attitude to the Chairman of the Information Technology Department at KAU for giving the full authorization and support to restructure and develop the Technical Communication course.

## REFERENCES

[1] Deanship of Admission & Registration, "Preparatory Year, King Abdulaziz University", 2016. Retrieved 16 February 2020, from https://admission.kau.edu.sa/Content-210-EN-260921

[2] FCIT - Faculty of Computing and Information Technology, "Summer Training Program CPIT-323 Syllabus", 2018. Retrieved 22 Jan 2020, from https://fcit.kau.edu.sa/aims/runReportAPI2.php?REP_ID=3&FL.

[3] R. AlGhamdi, "Fostering information technology students' internship program." *Education and Information Technologies* vol. 24, no. 5, pp. 2727-2739, 2019. doi:10.1007/s10639-019-09891-6

[4] S. Alkhalaf, S. Drew, R. AlGhamdi, and O. Alfarraj, "E-learning system on higher education institutions in KSA: Attitudes and perceptions of faculty members." *Procedia-Social and Behavioral Sciences,* vol. 47, pp. 1199-1205, 2012.

[5] M. Sabin, H. Alrumaih, J. Impagliazzo, et al., "Information Technology Curricula 2017. Association for Computing Machinery (ACM) and the IEEE Computer Society", 2017. doi: 10.1145/3173161.

[6] AlGhamdi R, Bahadad A (2018) Assessing the usages of LMS at KAU and proposing FORCE strategy for the diffusion. arXiv preprint arXiv:1902.00953

[7] F.W. Parkay, H.S. Beverly and D.G. Thomas. *Becoming a teacher*. Pearson/Merrill, 2010.

[8] H. Tompsett, H. Kathleen, M. B. Jane, G. M. Elaine, and T. Chris. "On the learning journey: what helps and hinders the development of social work students' core pre-placement skills?" *Social Work Education,* vol. 36, no. 1, pp. 6-25, 2017.

[9] C. McMorran, R. Kiruthika and L. Simei, "Assessment and learning without grades? Motivations and concerns with implementing gradeless learning in higher education." *Assessment & Evaluation in Higher Education.* vol. 42, no. 3, pp. 361-377, 2017.

[10] B. G. Trogden, and P. M. Joseph, "From Cornerstone to Capstone: Perspectives on Improving Student Communication Skills through Intentional Curricular Alignment." In *Communication in Chemistry*, pp. 17-28. American Chemical Society, 2019.

[11] Z.H. Duraku. "Factors influencing test anxiety among university students*". The European Journal of Social & Behavioural Sciences* vol. 18, no. 1, 2017.

[12] D. Baker and K.L. Gerald, *National differences, global similarities: World culture and the future of schooling*. Stanford University Press, 2005.

[13] KAU- King Abdulaziz University, "The Vice President for Educational Affairs sponsors the first introductory workshop for KAU Exit Exam", 2019. Retrieved 8 Feb 2020, from https://www.kau.edu.sa/Content-0-AR-276499

[14] D. Kennedy, Declan. *Writing and using learning outcomes: a practical guide*. University College Cork, 2006.

[15] KAU E-Learning Deanship, "Adopting the Blackboard system in the management of the educational process at the university", 2014. Retrieved 8 Feb 2020, from http://www.kau.edu.sa/Pages-Blackboard.aspx

[16] F. Vogt and M. Rogalla, "Developing adaptive teaching competency through coaching". *Teaching and Teacher Education*, vol. 25, no. 8, pp. 1051-1060, 2009.